\documentclass{emulateapj}
\usepackage[utf8]{inputenc}
\usepackage{color}
\usepackage{natbib}
\newcommand{\kep}{{\em Kepler}}

\newcommand{\eG}{{$\eta_{\textrm{Green Earth}}$}}
\newcommand{\jwst}{{\em JWST}}

\begin{document}

\title{SET-E: The Search for Extraterrestrial Environmentalism}
\author{Ben Montet\altaffilmark{1,2,3} and Ryan Loomis\altaffilmark{2,3}}
\date{April 1, 2016}

\email{btm@astro.caltech.edu}

\altaffiltext{1}{Cahill Center for Astronomy and Astrophysics, California Institute of Technology, 1200 E. California Blvd., MC 249-17, Pasadena, CA 91106, USA}
\altaffiltext{2}{Harvard-Smithsonian Center for Astrophysics, 60 Garden
Street, Cambridge, MA 02138, USA}
\altaffiltext{3}{But we did this on our own time, please don't blame our institutions.}

\begin{abstract}
    There is currently no evidence for life on any known exoplanet.
    Here, we propose a form of ``galactic anthropology'' to detect not only the existence
    of life on transiting exoplanets, but also the existence of environmental movements.
    By observing the planet's atmosphere over long time baselines, the destruction and
    recovery of a hole in an exoplanet's ozone layer may be observable. While not readily
    detectable
    for any one system with JWST, by binning together observations of hundreds 
    of systems we can finally determine the occurrence rate of environmental movements
    on Earthlike planets in the galaxy,
    a number we term \eG. 
\end{abstract}

\section{Introduction}

Now that the detection of planets orbiting other stars is pass\'e,\footnote{https://twitter.com/JasonFRowe/status/696792227629113344} we can begin searching these known planets for signatures of extrasolar life. The search has spanned the observational \citep{Wright14, Griffith15} to the theoretical \citep{Rauer11, Grenfell14}, making some predictions testable with current or soon-to-be-available
instruments \citep{Loeb07, Davenport13, Seager13, Snellen13} and others testable in the more distant future \citep{Loeb12, Loeb14, Stevens15}.

Should extraterrestrial life be discovered, there is no clear consensus as to whether humans should attempt to interact with it or not \citep{Serling62, Brin85}. While biosignatures will be detectable with upcoming telescopes such as the James Webb Space Telescope (JWST), it will be difficult to ascertain whether these signatures are the result of relatively simple organisms or of intelligent civilizations.

Given that we may wish to avoid interaction with malevolent or carelessly destructive lifeforms, it would be advantageous to find observable signatures of these traits. Atmospheric changes as a result of human interactions with the environment serve as a good starting template. Certainly, other species have significantly affected the environment on Earth: for example, plant life and phytoplankton have, over hundreds of millions of years, provided the Earth's atmosphere with a considerable quantity of oxygen. However, humans are the only known species to affect the atmosphere on a timescale of decades. Evidence of an exo-atmosphere changing on a similar timescale may then be evidence for intelligent life. 

Searching for signs of an advanced civilization destroying themselves and their environment is a rather grim endeavour \citep[e.g.][]{Stevens15}, so we propose instead to search for evidence of extraterrestrial environmentalism. In this brief paper, we describe how given high enough photometric precision and a long time baseline, it would be possible to observe the creation and diminution of a hole in the ozone layer of an exo-Earth. Such a hole could affect both the transmission signature of a planet and the transit light curve shape, and could be observed as seasonal variations in ozone abundance through reflected light spectroscopy.

\section{Atmospheric changes caused by (un)intelligent life}
    \subsection{The Earth's Ozone Hole}
        Ozone in the Earth's atmosphere is concentrated in the stratosphere, shielding life on Earth's surface from ultraviolet radiation \citep{Solomon99}. The mass production of chloro-fluoro-carbons (CFCs) in the mid-1900s, however, has caused substantial changes to the ozone layer on a relatively short timescale \citep[e.g.][]{Solomon88, Anderson91}. Thermal gradients in the atmosphere lead to the concentration of these ozone destroying molecules into seasonal clouds over the South Pole, severely depleting ozone and forming an `ozone hole' every spring \citep[e.g.][]{Crutzen86, Toon86, Shindell98}. Although global ozone concentrations have exhibited a long-term decline over the past decades, swift international action through the Montreal Protocol has significantly reduced CFC output and ozone levels are projected to recover in the coming century \citep{Elkins93, Weatherhead06, Newman06}. This has been widely accepted as one of the greatest successes of modern environmentalism.

\begin{figure*}[htbp!]
\centerline{\includegraphics[width=0.9\textwidth]{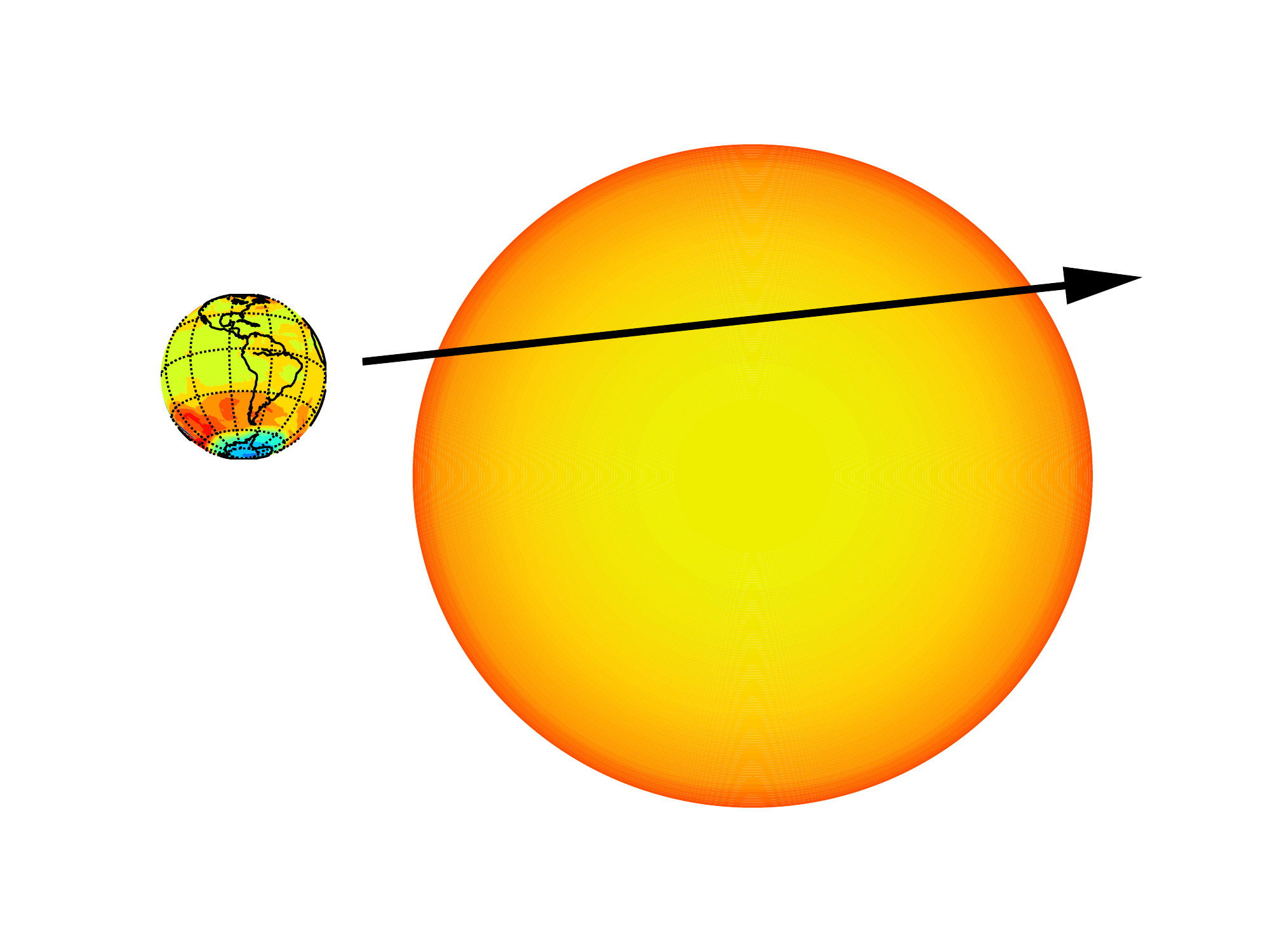}}
\caption{Simulated transit (not to scale) of an Earth analog passing in front of a Sunlike 
star. The Earth's ozone signature is highlighted, with bluer colors corresponding to 
positions on the Earth with less ozone and redder colors to positions with more ozone. 
Observing a transit at wavelengths sensitive to ozone (around $9.7\mu$m) might cause 
observable variations in the transit shape or time of ingress because of the non-uniform
distribution of atmospheric ozone. Ozone data from the Ozone Monitoring Instrument on the
NASA AURA spacecraft.}
\label{fig:Sensitivity}
\end{figure*}

    \subsection{Extraterrestrial Ozone Holes}
        Due to its radiation shielding abilities, the presence of ozone in an exoplanet atmosphere is likely valuable to carbon based lifeforms. If extraterrestrial intelligent life has the same propensity for ingenuity and self-destruction as humans do, however, they are likely to also find ways to damage their ozone layer. Very little research has been done on what such an `exo-ozone hole' might look like, but we speculate\footnote{Wildly} that there will likely be a long-term decline of total atmospheric ozone and seasonal variations similar to those seen on Earth. The expected latitudinal thermal gradient and transport of atmospheric particulates between atmospheric cells also suggests that ozone depleting clouds will likely form predominantly in polar regions.

\section{Observing Alien Ozone}

\subsection{Transmission Spectroscopy}

Assuming a favorable orbital geometry for the system, a planet will transit the face of its
host star once per year of the planet (Figure 1).

Ozone is detectable in the spectrum of the Earth in a $\sim$0.2 $\mu$m wide feature centered
at 9.7 $\mu$m \citep{Hovis68}.
This bandpass will be observable with MIRI on \jwst. 
Previously, this instrument has been considered by \citet{Barstow15} as a possible tool
to detect ozone features, however they only considered detecting the existence of ozone, not searching for long-term
variations. By stacking together many spectral observations, these authors find that 
ozone will be detectable in the atmosphere of an Earth-like planet orbiting an M dwarf 
using only observations from 30 transits with \jwst.

We propose observing transits with both \jwst\ and its successor. Given a long time baseline and enough transits,
one could then fit a variable ozone model (such as a long-term trend) to the observations and infer the rate of change of ozone in the atmosphere. Here, the most important consideration is the long time baseline required.
Since \jwst\ will have a short lifespan relative to the timescale for the 
evolution of our own ozone layer, future mid-IR space observatories will be essential 
to complete this study.

\subsection{Transit Morphology}

The signature of an ozone hole, rather than a uniform decrement in the atmospheric ozone,
could also be observable in changes in the transit shape as a function of wavelength. 
The Earth is largely spherical \citep{Eratosthenesbc}, and typical transit models 
reflect this \citep{Mandel02}. However, by observing in a narrow wavelength range, centered on the region of the spectrum
affected by ozone, the Earth would be asymmetric: it would have a smaller effective radius
at the latitudes corresponding to the ozone hole and a larger effective radius where there 
is no ozone hole, making the silhouette of the Earth slightly egg-shaped.

Since atmpospheric ozone is typically located in the stratosphere approximately 20 km above the surface
of the Earth \citep{Shindell98}, a hole covering $20\%$ of the surface would affect the
observed transit depth by up to 0.1 parts per million, a roughly 0.1\% effect (Figure 2). The hole
could also affect the time of ingress/egress by up to
0.6 seconds, depending on the orientation of the transiting planet with respect to its host 
star.
Through observations with very high resolution photometry and very fine time sampling, 
taken simultaneously in a bandpass focused on ozone and one at a different wavelength, this 
could be trivial to detect.\footnote{We leave this exercise to the reader.}

\begin{figure}[htbp!]
\centerline{\includegraphics[width=0.40\textwidth, angle=270]{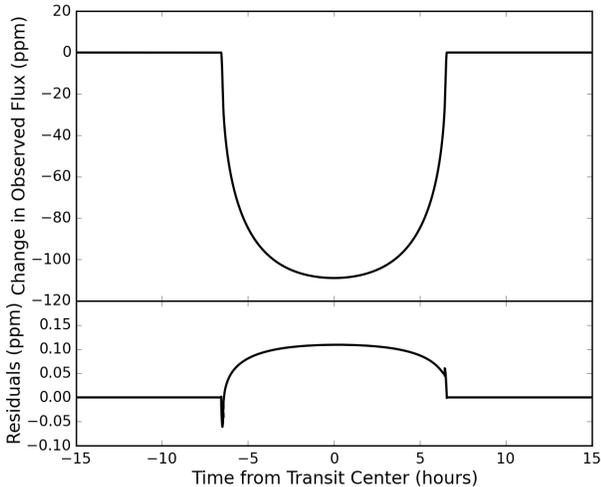}}
\caption{(Top) Simulated transit of an Earthlike planet transiting a Sunlike star, with
a full ozone layer (solid line). This transit is significantly affected by removing the
ozone layer on the bottom $20\%$ of the Earth's atmosphere, effectively decreasing the observed
radius of the planet in that region by 20 km, although the difference cannot be seen
on this scale. (Bottom) The residuals between these two models, considering an alignment
where the ozone hole is aligned such that it is pointed in the direction of motion of Earth.
The difference at first and third contact are due to this hole's geometry, where the overall
change in transit depth is observed at all times. This effect may be small, but for a larger planet or a smaller star, the signal can grow arbitrarily large.
}
\label{fig:Sensitivity}
\end{figure}

\subsection{Ozone in Reflected Light}
While variations in ozone in transit may be hard to detect with current\footnote{Or future}
facilities for any single system, we might have more hope in detecting seasonal variations
in ozone over the course of the planet's orbit.
This is already a well-established problem. 
As the Earth's orbital speed around the sun is $30$ km s$^{-1}$, by observing the planet
at various points during its orbit, one can separate the planetary spectrum from the
stellar spectrum by searching for changes in the wavelengths of spectral features
\citep[e.g.][]{Snellen14, Brogi16}.
Since the seasonal variations in ozone can be 25\% or more \citep{Cariolle90}, observing
throughout an orbit could be useful in detecting environmental movements.

\section{Demographics of Ozone Holes in the Galaxy}

As previously noted, \citet{Barstow15} show that ozone features could be detected by binning
together multiple transits observed with \jwst.
However, even ignoring the relatively short time baseline offered by 
\jwst, observing changes in the ozone layer for any particular system may be
difficult: a 4\% change in the total ozone would only represent a 0.1$\sigma$ effect
in the observed ozone signature.

Perhaps we should, instead of focusing on any one system, try to understand the population
of ozone holes in the galaxy. 
Very small effects can be understood on a population scale by binning together observations
of many similar objects. 
For example, \citet{Sheets14} combined secondary eclipses of sub-Saturn planets in \kep\ data
to estimate the average albedo of these planets, while \citet{Zuluaga15} combine transit
photometry of giant planets in a statistical search for exo-rings.

We can do the same to better understand the underlying statistics of ozone holes.
Instead of analyzing any single system, we can instead observe many transits of $\sim 5000$ 
Earth-like planets over decades. If every one of these planets had an ozone hole forming,
then we could easily observe these variations as a $7\sigma$ effect in the binned transmission
spectrum.
Depending on what fraction of these planets have life that is affecting their ozone layer,
this number would be smaller. 
By observing these systems across several decades, the actual number of environmental
movements currently ongoing in the solar neighborhood, \eG, can be inferred.
By replacing $\eta_\oplus$ in the Drake Equation with \eG, this form of
galactic anthropology allows us to recover not only the
occurrence rate of advanced civilizations in the galaxy, but the occurrence rate of 
civilizations that we might actually want to contact.

\section{Discussion}

    \subsection{Feasibility of observations}
        The observations proposed in this paper are likely not feasible, and even if they were, the transient nature of the feature makes it incredibly unlikely to be observed. However, it is possible to test this avenue of thought by observing the ozone layer in the Earth using spaceborne spectrometers such as CIRS on Cassini. If it were to take a short break from doing actual useful science during a transit of the Earth in front of the Sun, and instead attempted to observe the ozone signature in the Earth's atmosphere, it would showcase the potential of this method.
        
        Additionally, any observations searching for exo-ozone holes will likely have to be conducted from space. The Earth's atmosphere is not transparent (yet) at $9.6 \mu$m because of our own ozone, but by moving to space this feature of the atmosphere can be avoided. As we may not want to wait until all ozone in the Earth's atmosphere has been
        depleted, we could just use \jwst, which is scheduled to be launched soon \citep{Zackrisson11}, as well as its eventual successors.

    \subsection{Possible complications}
        Assuming a depletion in alien ozone is dominant at one pole, much like our own, this hole would only be observable in transit for favorable orientations: depending on the planet's axial tilt and the scale height of the atmosphere, it is possible that the hole would be fully hidden by the shadow of the planet, leaving it unobservable to an Earthbound observer. Even if the planet has a unfavorable axial tilt, however, we should always be able to detect long-term evolution in the total atmospheric ozone concentration through reflected light spectroscopy.
    
        In the unlikely situation that variations in exo-planetary ozone levels are detected, their interpretation might be complicated by a number of factors. Atmospheric ozone is affected by cosmic ray flux \citep{Lu01, Tabataba16}, so in order to interpret the observations we must also observe and understand the activity of the host stars. A mission like \kep, which was built to measure precise transit photometry, but with an additional photometric filter centered on the wavelengths most sensitive to ozone in the mid-infrared, would be ideal for such a purpose. 
        The original \kep\ mission cost approximately $600$ million dollars, but given that we already \textit{have} a \kep, this time we could save significantly on research and development costs (except, perhaps, for the development of sturdier reaction wheels).

        Observations of ozone could also be confused by high levels of volcanic activity \citep{Prather92} or high levels of thermonuclear warfare on the planet in question \citep{Stevens15}. However, given the sensitivity required to detect extrasolar environmental movements, either of these would likely create time variable infrared features that would be easily distinguishable from the signal considered here.

    \subsection{Future implications}
        If changes in an exo-ozone hole are detected, this would open up additional avenues for galactic anthropology. The timescale between the initial decrement in ozone and the eventual return to normalcy could provide insights into the level of scientific funding as well as the scientific literacy of the political leaders on that planet.\footnote{As opposed to on this planet: http://gawker.com/donald-trump-decries-war-on-hairspray-1750373974}

\acknowledgements
    Thanks to Ian Czekala for reading a draft of this manuscript and providing valuable feedback.
    We also thank both Sarah's Market and Caf\'e and Armando's Pizza and Subs,
    both of Cambridage, MA, for the beer and sandwiches which enabled much of the work in 
    preparing this manuscript.
    We would especially like to acknowledge the ozone layer on the Earth for shielding us 
    from UV radiation throughout the preparation of this manuscript.

\end{document}